\documentclass[11pt]{article}
\usepackage{array}
\usepackage{graphicx}
\usepackage[noadjust]{cite}
\usepackage{textpos}
\usepackage{bbold}
\usepackage{adjustbox}

\title{QCD at 50: Golden Anniversary, Golden Insights, Golden Opportunities
\footnote{Invited keynote lecture at the 2023 Erice Summer School.  This write-up incorporates some corrections and improvements from the talk as delivered.   I presented versions of this material at the 2023 Quantum Connections summer school in Stockholm, the ``50 Years of QCD'' conference at UCLA, at MIT, and as parts of lecture series at SJTU in Shanghai and most recently at ASU in Tempe.}
\author {Frank Wilczek  \\ \\
\small\it Center for Theoretical Physics, MIT, Cambridge, MA 02139 USA; \\
\small\it T. D. Lee Institute and Wilczek Quantum Center, \\
\small\it Shanghai Jiao Tong University, Shanghai, China;\\
\small\it Arizona State University, Tempe, AZ, USA; \\
\small\it Stockholm University, Stockholm, Sweden }}

\begin{document}

\maketitle

\begin{textblock*}{5cm}(11cm,-8.2cm)
\fbox{\footnotesize MIT-CTP/5689}
\end{textblock*}

\begin{abstract}
The bulk of this paper centers around the tension between confinement and freedom in QCD.  I discuss how it can be understood heuristically as a manifestation of self-adhesive glue and how it fits within the larger contexts of energy-time uncertainty  and {\it real virtuality}.  I discuss the possible emergence of {\it treeons\/} as a tangible ingredient of (at least) pure gluon $SU(3)$.  I propose {\it flux channeling\/} as a method to address that and allied questions about triality flux numerically, and indicate how to implement it for electric and magnetic flux in material systems.  That bulk is framed with broad-stroke, necessarily selective sketches of the past and possible future of strong interaction physics.  At the end, I've added an expression  of gratitude for my formative experience at the Erice school, in 1973.  

\end{abstract}

\medskip

\bigskip
\bigskip

\begin{figure}[hbt!]
    \centering
    \includegraphics[width=\columnwidth]{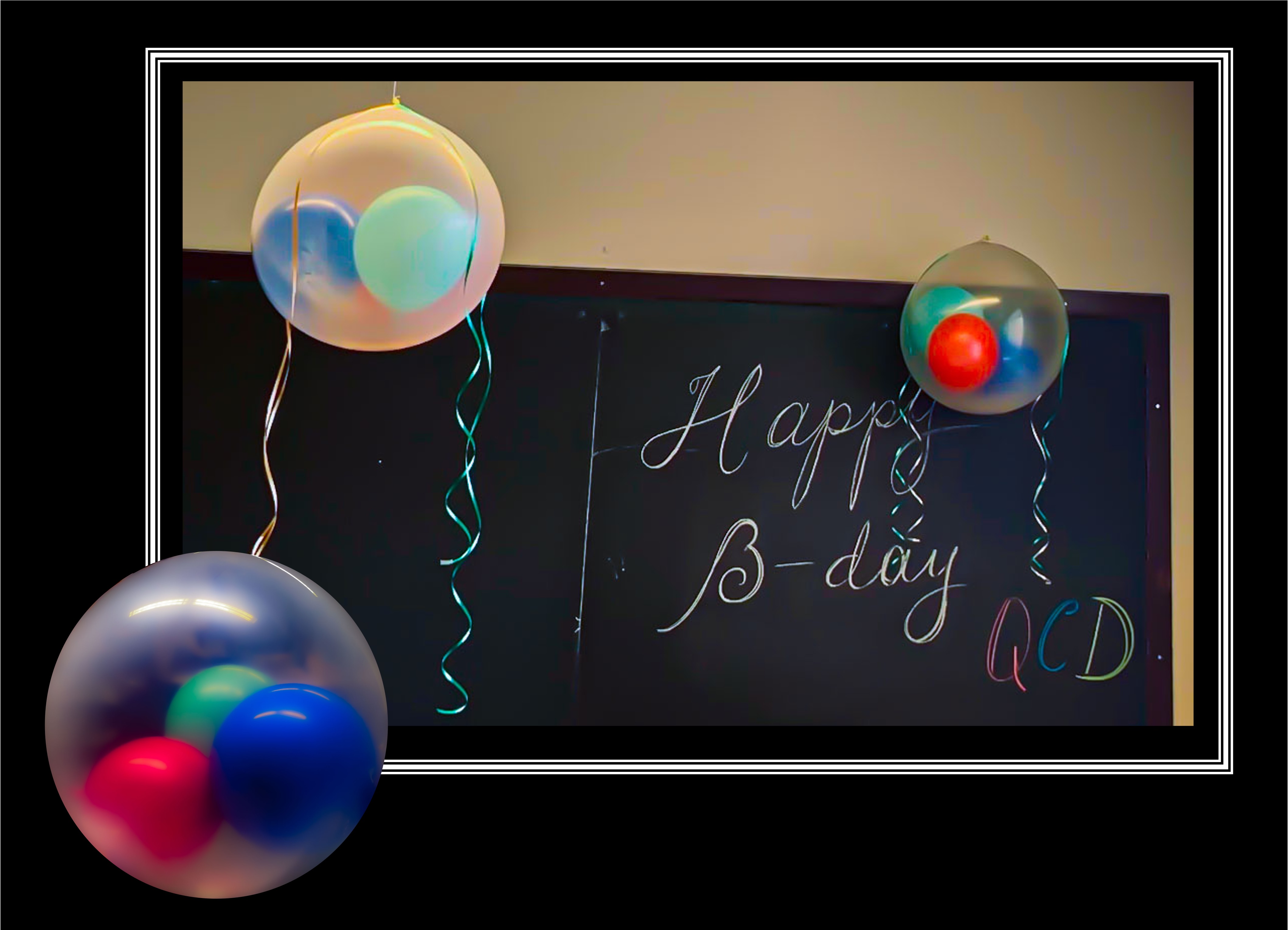} 
    \caption{Photo from a small surprise party in my office at MIT, April 2023, featuring remarkable feats of confinement.}
     \label{qcd_balloons} 
\end{figure}

It is usually straightforward to assign unique dates of birth to people, but for ideas it is rarely so.  That said, I think the organizers' choice to date 50$^{\rm th}$ anniversary of QCD to 1973, marking the appearance of the basic papers on asymptotic freedom and the theory of the strong interaction it implies (\cite{gw1}-\cite{gp}; see also \cite{tH1, gM , aps_collection}), is not unreasonable.  In any case,  I'm happy to run with it.  (See Figure \ref{qcd_balloons}.) 

\section{A Brief Chronicle of the Strong Interaction}

As seems appropriate for the occasion, I will frame the more technical body of this talk with brief perspectives on the pre-history and the future of QCD.   I call this opening a ``chronicle'' rather than a ``history'', because it highlights only a very small number of intellectual milestones, and totally neglects the details, the accumulated incremental progress  - and the missteps -- that characterized the actual events in real time.  Here is my chronicle:

\begin{itemize}
\item Geiger-Marsden experiment and the Rutherford-Bohr model: {\it Atoms have nuclei, and a new force is discovered.}  

In this ``deeply elastic'' scattering experiment, familiar to all physicists, the occurrence of large deflection angles when $\alpha$ particles impacted a thin gold foil showed that atoms are capable of ``hard'' scattering, i.e. large momentum transfers.   Rutherford both suggested the experiment and then interpreted it quantitatively, as revealing that gold atoms contain tiny central nuclei wherein large amounts of positive charge and essentially all the atom's mass are concentrated.  Given that assumption, the angular dependence of the scattering cross-section could be understood quantitatively based on Coulomb repulsion.  

Soon afterwards, Bohr introduced rules involving Planck's constant that succeeded in accounting, on the basis of Rutherford's atomic model (nucleus surrounded by electrons), for the existence of atomic spectra -- quantitatively for hydrogen, and qualitatively in general.  This success made it plausible that all of atomic physics, and even chemistry, might be accessible to theoretical interpretation based on electrodynamics and a future, mature version of quantum theory.   

The existence of nuclei, however, indicated that there had to be another force, more powerful than electromagnetism, capable of binding lots of positive charge within a small volume.  This mysterious new force was called the strong force.  Pre-1913 physics had given no hint of  the existence of that force.  Something new was needed, but exactly what was mysterious.  

\item Existence of neutrons (Chadwick): {\it Classical nuclear physics -- an extremely fruitful model!}  

James Chadwick's discovery of the neutron made it possible to construct a semi-quantitative account of atomic nuclei as collections of protons and neutrons.  (Ultimately, this is possible because nuclear binding energies are of order tens of MeV, which is significantly less than the rest mass of protons and neutrons and also the gap separating them from excited baryons.)    Modeling based on that foundation -- ``classical'' nuclear physics -- was, and remains, a tremendously useful enterprise.  It has given us, among other things, penetrating insight into how stars shine and evolve, a richly detailed account of origins and abundances of nuclear species, and important medical and energy technologies.

\item Resonances and the quark model: {\it The ``old'' QCD correlates lots of surprising data.}

Ironically, however, classical nuclear physics did not point the way to a profound microscopic theory of the strong interaction.  (And conversely, as yet QCD has had only marginal impact on classical nuclear physics.   Addressing that disappointing shortfall  is an  important challenge for future research, as I'll highlight below.)  Early hope that the problem of the strong interaction would be solved within the ``neo-Newtonian'' framework of forces between protons and neutrons that would be determined along the lines 
$$
{\rm Geiger-Marsden \ experiment \ \ + \ \ Rutherford \ interpretation } \\  
$$
$$
\Rightarrow \ \  {\rm Coulomb \ force \  law}
$$
did not long survive experimental scrutiny.  Rather than merely deflections, collisions between hadrons frequently gave birth to many particles, with energy converted into mass.  Many new and previously unsuspected unstable particles, sometimes seen only as ``resonance'' enhancements  of carefully processed exclusive cross-sections, were discovered.  Flavor $SU(3)$ and the quark model organized and somewhat tamed that gush of data.  The quark model, however, was only semi-quantitative, and it invoked some very peculiar rules.  In this it resembled Bohr's atomic model, but without  analogs of that model's showcase hydrogen atom, applicable (electrodynamic) force laws --  or evidence for electrons!

\item Deep scattering: {\it Snapshots of protons expose surprising simplicity.}

The epochal deep inelastic scattering experiments at SLAC \cite{slac_expts},  starting in the late 1960s, provided snapshots of proton interiors.  Those experiments can be seen -- clearly, in retrospect -- as a logical continuation of the Geiger-Marsden experiment \cite{bj}\cite{partons}.  But in order to do justice to proton interiors, which are small and where things move fast, one must bring in virtual photons that have very large transfers of energy as well as momentum,  in order to achieve sufficient temporal and spatial resolution.   The resulting snapshots revealed that particles with properties resembling those expected for quarks really do exist, but also that  they do not exhaust the proton's content. 

\item The new QCD emerges: {\it Asymptotic freedom guides us to specific equations and sharp predictions.} 

It was difficult to accommodate the paradoxical behavior of quarks  inside protons -- that they move freely most of the time, yet remain permanently confined --  within the framework of quantum field theory.  The difficulty of that challenge proved to be a wonderful blessing in disguise. As David Gross and I, and independently David Politzer, discovered,  non-abelian gauge theories -- and they alone -- are uniquely suited to do  the job.    That discovery enabled us to credibly propose a unique, specific set of equations for the strong interaction: the new QCD, or simply QCD, whose golden anniversary we celebrate. 

\item Fifty years of survival and growth: {\it From triumph to service ... with no end in sight}.

While the deep scattering experiments were inspiring and suggestive, several steps of interpretation separated the raw observations from their theoretical interpretation in QCD, and initially both sides of the comparison were imprecise.  So at first there was plenty of room for skepticism.  

The new QCD got a big boost from the ``November Revolution'' of 1974 \cite{ting}\cite{psi}, which revealed a spectrum of mesons that unmistakably resembled the positronium spectrum \cite{ap}, and begged for an interpretation as weakly coupled heavy (charm) quarks and antiquarks.  It was not entirely straightforward, but over time people learned what measurements to make, and what calculations to do, in order to test the theory rigorously, and then to put it to work.  Experiments at higher-energy accelerators showed clearly defined jets.  Thus, if you squint, you can see the Feynman diagrams of QCD laid out before you.  As both experimental technique and calculational power improved the focus of the high-energy physics community, reflected in the programs of big international conferences, turned  from the exploratory excitement of {\it testing QCD\/} to the hard but useful and reliably rewarding work of {\it calculating backgrounds}.  

\end{itemize}

The elucidation of the strong interaction ranks high among the most impressive and comprehensive triumphs in the history of science, I think.   It is also a gift that I expect will keep on giving, as I'll expand upon at the end. 

\bigskip

\bigskip

The next several sections, up to the last two, give perspectives of what is perhaps the most unusual and characteristic feature of the strong interaction (and QCD), that is its complementary aspects of confinement and freedom.  This part could be called ``Confinement Made Simple, Then Complicated''.

\bigskip

\bigskip

\section{Confinement and Freedom}

The problem of confinement was a theoretical discovery, seen at first as a challenge to experiment.  It first arose in the old QCD, with the weird rules of the quark model.   The known hadrons were built following two body plans: the mesons, with a quark and an antiquark, and the baryons (or anti-baryons) with three quarks (or antiquarks).  Despite vigorous searches, and several false alarms, isolated quarks were never observed.   

The introduction of a color degree of freedom for quarks \cite{greenberg} enhanced the respectability of the model, by allowing quarks to satisfy a normal spin-statistics connection.  It also allowed a more cogent statement of the confinement hypothesis, as the claim that color singlet states are systematically lighter than non-singlets.  In a visionary paper \cite{nambu} Nambu even connected that observation to the possibility that the color charge is associated with an $SU(3)$ gauge theory.   

\begin{figure}[hbt!]
    \centering
    \includegraphics[width=\columnwidth]{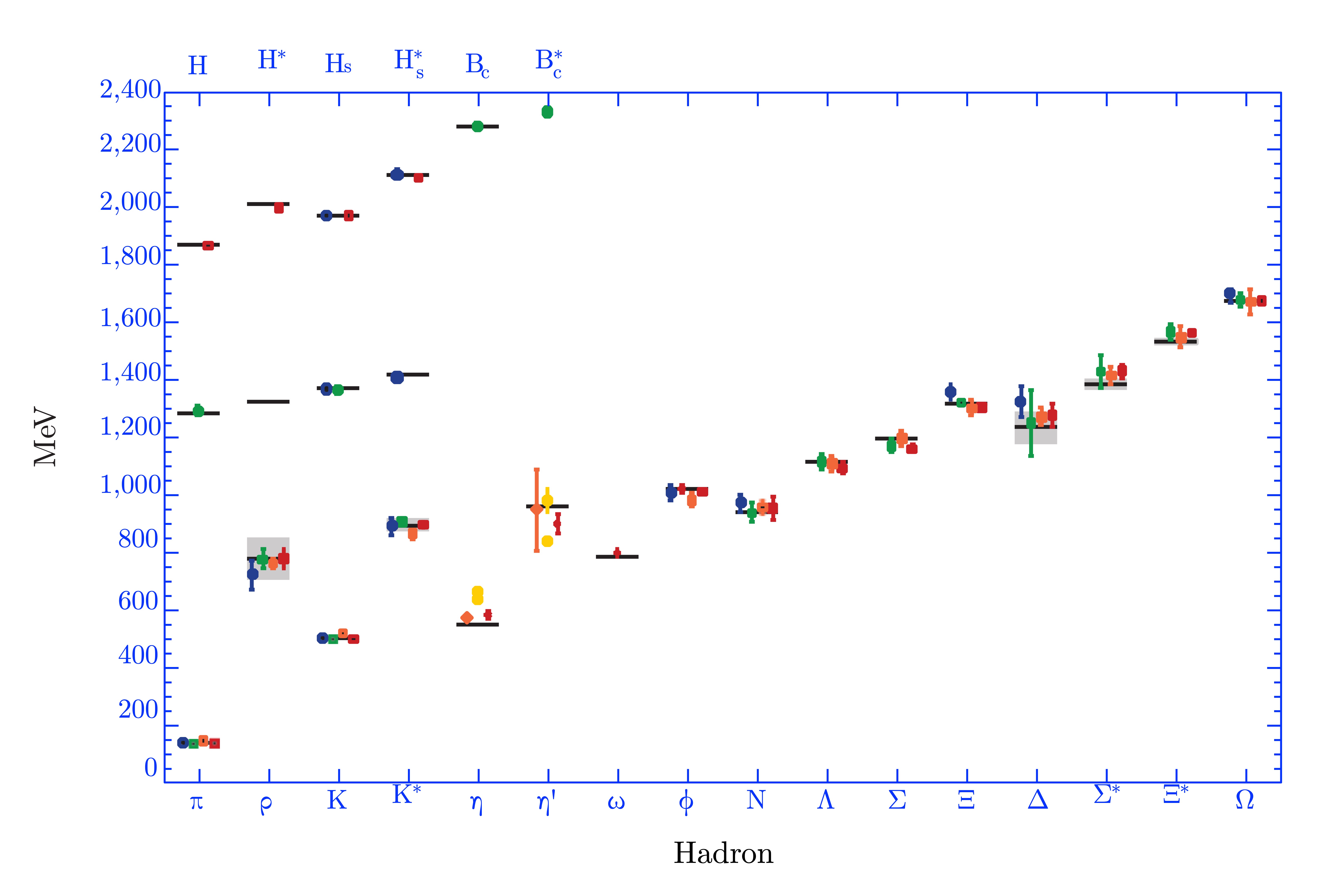} 
    \caption{Comparison of calculated hadron masses with experimental data, for the low-lying spectrum and also (upper left) representative heavy quark systems.  The points with their error bars are calculations; the lines or, for a few broad resonances, rectangular clouds are the measured values.  The calculation of the low-lying spectrum has only three continuous parameters as input: $m_u + m_d$, $m_s$, and the gauge coupling.  Moreover, a calculable combination of those simply sets the overall mass scale.  The correct spins and other quantum numbers are reproduced, and no extra states appear.  From \cite{kronfeld} }
\label{hadron_masses} 
\end{figure}

In the new QCD confinement became, in principle, a problem that could be -- and had to be -- addressed theoretically.  After all, it is part of calculating the physical spectrum!   Impressive heuristic arguments for the basic phenomenon, in the sharp form that only color singlets appear  in the spectrum, appeared almost immediately \cite{wilson} \cite{susskind}, followed shortly by convincing numerical evidence \cite{creutz}.  Today, after decades of heroic work from the lattice gauge theory community, we have calculations of the low-lying spectrum that agree in full detail with experimental measurements, at the few per cent level. (See Figure \ref{hadron_masses}.)  

\begin{figure}[hbt!]
    \centering
    \includegraphics[width=\columnwidth]{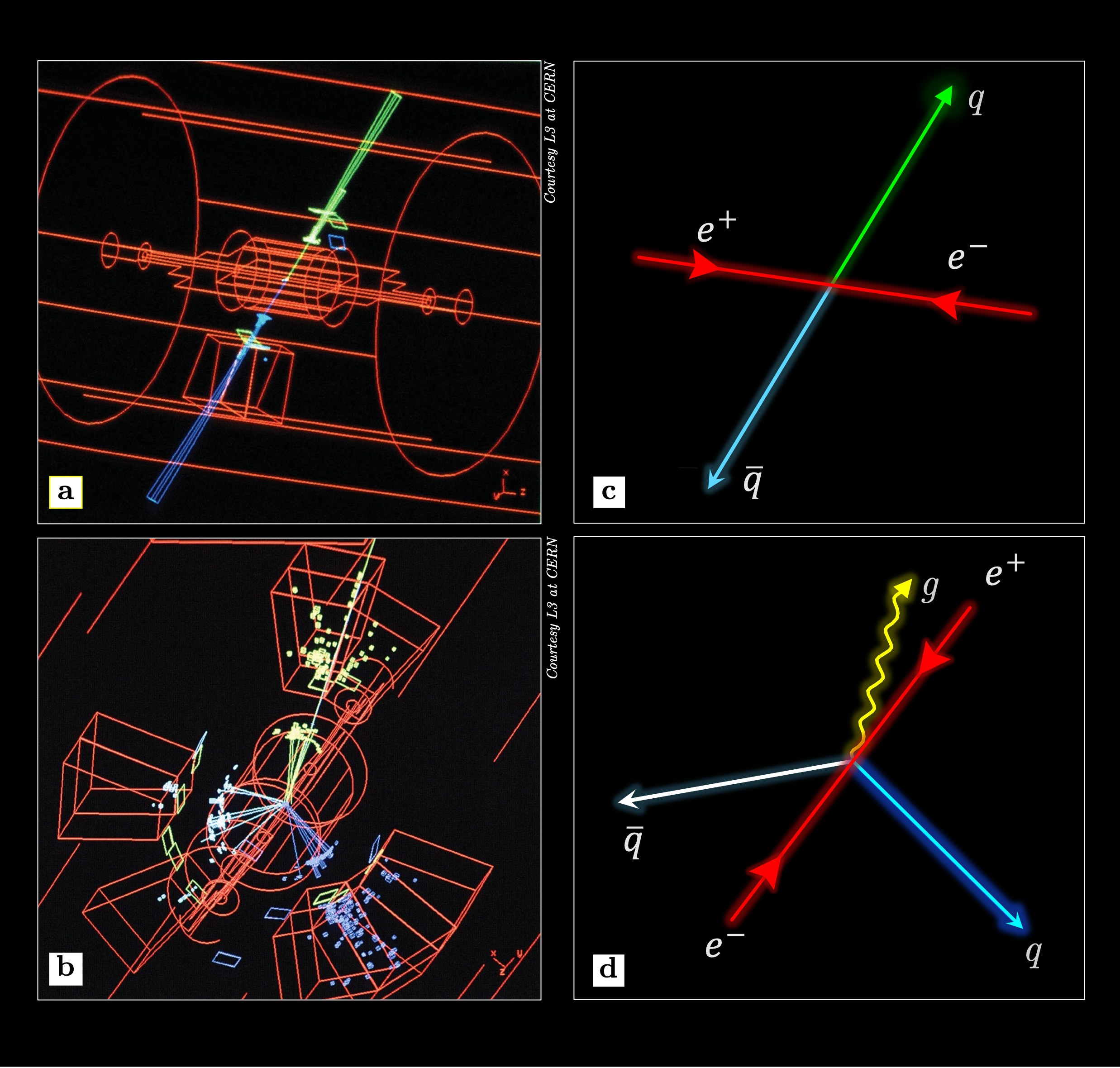} 
    \caption{At high energy and momentum transfers, one observes that the energy-momentum flows of hadrons have a jet-like structure. As observed in many experiments, the distribution of such jets in number and overall energy-momentum match perturbative calculations identifying them with individual quarks, antiquarks, and gluons. {\bf a, b:}
Electron-positron collisions at LEP, from the L3 collaboration. The alignment of energetic particles in jets is visible to the naked eye.
{\bf c:} QCD interpretation of a:   Electron and positron annihilate into a virtual photon which materializes into a quark-antiquark pair. The quark and antiquark usually dress themselves with soft radiation, and we observe a two-jet event. {\bf d:} QCD interpretation of b:  Roughly one-tenth as often, however, a hard gluon is radiated. Then quark, antiquark, and gluon all dress themselves with soft radiation, and we see three jets.}
\label{jets} 
\end{figure}

The ``freedom'' of quarks was an experimental discovery, seen at first as a challenge to theory.  As mentioned above, deep scattering experiments suggested the existence of quarks that propagate approximately freely for short times and distances within protons.  That behavior was, from the perspective of relativistic quantum field theory, deeply problematic.  The discovery of asymptotic freedom, occurring within a unique non-abelian gauge theory that could plausibly represent the strong interaction -- i.e., the new QCD --  answered that challenge.  Today people can, and routinely do, ``see'' quarks and gluon propagation quite vividly.  (See Figure \ref{jets}.)

Thus, both confinement and freedom are well-documented properties of QCD, and of the observed strong interaction.   Yet there remains undeniable intellectual tension between them.  The Clay foundation offers a big prize for mathematical proof of (a weak form) of confinement \cite{clay}, and as yet no one has claimed it successfully.  For our peace of mind, and with the hope that better understanding will have fruitful spin-offs, we must face up to this tension.

\section{An Instructive Model}

3+1 dimensional non-abelian gauge theories are gloriously complex.  Following Einstein's dictum to ``make things as simple as possible, but no simpler'', it proves good strategy to approach our problem through a much simpler theory, namely 1+1 dimensional abelian gauge theory.  (Schwinger \cite{schwinger} famously pioneered the study of this model, and noted the drastic difference between the physical spectrum and the underlying degrees of freedom.  Among many later related studies let me mention the widely used  phenomenological Lund model \cite{lund}; and \cite{coleman} for considerations close to those that follow here.)  Specifically, let us consider the theory of a spin-1/2 ``Quark'' with change $q$ and mass $m$ minimally coupled to a Maxwell gauge field.  Using units with $\hbar = c = 1$ and standard relativistic notation, the Lagrangian density is:
\begin{equation}
 {\cal L} ~=~ -\frac{1}{4}F_{\mu \nu} F^{\mu \nu} \, + \, \bar Q \bigl( \, \gamma^\mu (i \partial_\mu - q A_\mu ) + m \bigr) \, Q
 \end{equation}
 
Note that in 1+1 dimensions charge has mass dimension unity, $[q] = M$.  We will be interested in the weak-coupling limit $q/m << 1$.  As we shall see, in that limit the theory becomes very simple indeed (for a relativistic quantum field theory). Yet it is not too simple to exhibit both confinement and freedom.   
 
In the gauge $A_1 = 0$, the Maxwell term contains no time derivatives, so the gauge field is enslaved to the Quark,  The one-dimensional form of Gauss' theorem tells us that the field between a static Quark and anti-Quark is constant, and therefore so is the energy density.   Thus we have a linear potential, and a constant force.  In our weak-coupling regime the resulting acceleration is small, so the quasi-static approximation is appropriate as well as tractable.  

Though the force is weak, it is inexorable.  The weakness of the force implies approximate freedom for the Quarks; its inexorability implies ultimate confinement.  Since the flux emanating from an isolated Quark cannot terminate, an isolated Quark is associated with infinite energy, and cannot be produced.  

\section{A Paradoxical S-Matrix}

Thus, this model embodies both freedom and confinement.  Since it does so in a simple, physically transparent way, we can use it to illuminate the conceptual tension between those two behaviors.  

As a concrete and quasi-realistic thought-experiment, let us consider an analogue to $e^+ e^-$ annihilation in this model.  Thus, we introduce an additional, ``probe'' $U(1)$ gauge field, ultra-weakly coupled to ``leptons'' $E$ and also to our ``quarks'' $Q$.  When we annihilate $E^+E^-$ pairs, the virtual Photons can communicate with the Quarks.  What happens?

On the one hand, it seems intuitively obvious that following creation of a Quark-antiQuark pair at a point with a decent amount of spare kinetic energy they will propagate away almost freely for a long time, since the attractive force is feeble.  

On the other hand, consider the $S$-matrix.   Again, since the coupling is weak, we can expect that it is valid, at a first go, to approach the problem using the non-relativistic Schrödinger equation and a linear potential.  This gives us a discrete spectrum.  For large quantum numbers (many nodes) we can use the WKB approximation, to determine the energies and spacings, in a few strokes:
\begin{equation}
2\pi n ~=~ \int dx\, p ~=~ 4 \sqrt{2m} \int\limits^{\frac{2E_n} {Q^2}}_0 \, dx \, \sqrt{E_n - \frac{Q^2}{2}x} 
\end{equation}
\begin{equation}\label{meson_spectrum}
E(n) \, \propto \, n^{\frac{2}{3}} \frac{Q^\frac{4}{3}}{m^{\frac{1}{3}}}
\end{equation}
\begin{equation}
\Delta E(n) \, \propto \, n^{-\frac{1}{3}} \frac{Q^\frac{4}{3}}{m^{\frac{1}{3}}}
\end{equation}
Of course, we should add in the rest mass $2m$.  

Above the two-meson threshold, $E = 4m$ (plus a little) we have the possibility of decay or, alternatively viewed, two-meson production, essentially through a string-breaking process.  Also, we have the possibility of annihilation back into leptons, through the ultra-weak probe interaction.  But neither these nor other refinements seriously affect the broad conclusion that over a wide swathe of energies, including a broad range where we might expect nearly free propagation, there are gaps  in the spectrum.  Within those gaps, according to the S-matrix, nothing happens!  Here we confront, in a sharp and concrete form, the tension between freedom and confinement. 

\section{Reconciliation: Eternal and Transient}

\subsection{Time-Energy Uncertainty}

The resolution of this paradox is profoundly simple: {\it The spectrum and the S-matrix reflect behavior over infinitely long times - but life is finite}!   

Let us ground that insight in instructive equations and procedures.  The concept of time-energy uncertainty has a mixed reputation.  There is clearly something right about it, and it is commonly invoked as a source of qualitative insight.  On the other hand, because there is no  quantum-mechanical operator that corresponds to the coordinate $t$, one cannot carry over the usual approach to position-momentum uncertainty, epitomized in a simple, general equation.   In reality, the situation for time-energy is not as murky as it is usually made to appear \cite{baez} -- nor is position-momentum as clear.   Indeed, as befits special relativity, they are both on the same footing, as we'll now discuss.

A very general uncertainty principle follows directly from any non-zero commutator 
\begin{equation}\label{c_r}
[ X, Y ] = i Z
\end{equation}
involving three Hermitian operators $X, Y, Z$.  We will be evaluating these operators in a particular state, and to obtain the sharpest result we should remove the expectation values from $X$ and $Y$, so $X - \langle X \rangle \leftarrow X, Y - \langle Y \rangle \leftarrow Y$.  That subtraction does not change the form of the commutator Eqn.\,(\ref{c_r}).  For any real number $\lambda$ we have
\begin{equation}
0 \leq \langle (X-i\lambda Y)(X + i\lambda Y)\rangle
= \langle X^2\rangle + \lambda^2 \langle Y^2\rangle  + \lambda \langle Z \rangle
\end{equation}
since this is the norm of $(X + i \lambda Y) \rangle$.  Thus this quadratic expression in $\lambda$ cannot have non-degenerate real roots, and we deduce
\begin{equation}
\langle Z\rangle^2 \leq 4 \langle X^2\rangle \langle Y^2\rangle
\end{equation}

Applying this general uncertainty principle to the dynamical equation for operators (in the Heisenberg picture)
\begin{equation}
[ H, A ] = i \frac{dA}{dt}
\end{equation}
gives us
\begin{equation}\label{operator_uncertainty}
\frac{1}{2} \leq \frac {\langle (\Delta E )^2 \rangle^{1/2}\  \langle (\Delta A^2) \rangle ^{1/2}} {|\langle \frac{dA}{dt} \rangle |}
\end{equation}

This is an uncertainty principle that brings in time and energy.  It also brings in an operator $A$.  If we want to interpret the reading of $A$ as a measure of time, we want $A$ to be a clock, such that $A$ stands in for $t$ in the state of interest.  If we simply substitute $t$ for $A$ in Eqn.\,(\ref{operator_uncertainty}) we get the ``naive'' form $\frac{1}{2} \leq {\langle (\Delta E )^2 \rangle^{1/2}\  \langle (\Delta t^2) \rangle ^{1/2}}$.  Because $t$ is not an operator there is no universal Platonic ideal clock; but we can and do aspire to design good clocks adapted to particular classes of states.  

It is worth remarking that in relativistic quantum field theory $x$ is no more of an operator than is $t$.  (Energy and momentum density, on the other hand, are local observables.)   As we have just seen for time, if we want to measure ``position'' we must specify an observable adapted to the state in question; and if we want to quantify its uncertainty, then strictly speaking we must use a relation parallel to Eqn.\,(\ref{operator_uncertainty}).     In all cases, suitable operators constructed from products of fields primarily supported in distinctive time-space regions can provide signals of activity in the regions where they are supported.  

Returning to our paradoxical S-matrix, note that the meson spectrum of Eqn.\,(\ref{meson_spectrum}) becomes very closely spaced as the number of nodes $n$ gets large, according to
$$
\Delta E(n) \, \propto \, n^{-\frac{1}{3}} \frac{Q^\frac{4}{3}}{m^{\frac{1}{3}}}
$$
It is for large $n$, of course, that semiclassical dynamics should be valid.   But in that regime the energy levels become very closely spaced, and energy-time uncertainty implies that they take a long time to resolve.  Likewise, it takes a long time for the  S-matrix to become a good approximate description of dynamics.   

Thus, our paradox is an instance of complementarity: different, mutually exclusive concepts supply answers to different questions about the same system.   If we want an accurate account of energy, then the spectrum and S-Matrix are appropriate tools; if we want an accurate account of motion and change, then ``time'' - creatively defined - is the appropriate tool.

\subsection{Real Virtuality}

\begin{figure}[hbt!]
    \centering
    \includegraphics[width=\columnwidth]{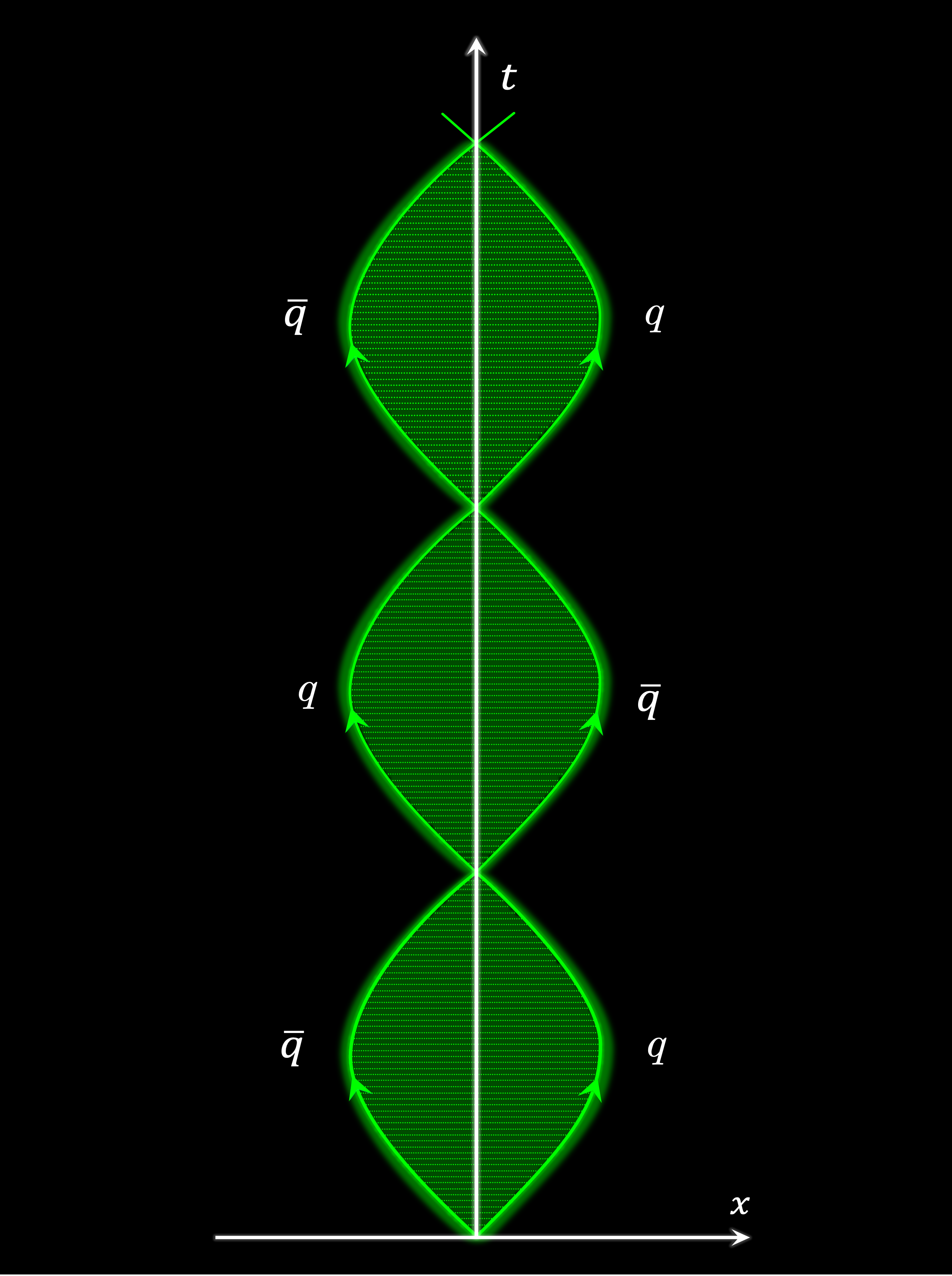} 
    \caption{Space-time diagram showing the world-lines of a fictitious weakly coupled Quark-anti-Quark pair following their creation with non-zero momenta at a point. Repeated cycling supports the possibility of temporal interference.}
\label{world_loops} 
\end{figure}

It is entertaining to consider how the semiclassical trajectory picture goes over into the discrete spectrum and gappy S-matrix as we shed time resolution.  Our ``quarks'' ,  once created significantly above threshold, at first separate almost freely, since the force between them is weak.  But as they separate they are doing work, and eventually they run out of kinetic energy and come back together.  Then their world-lines cross again, now with equal and opposite velocities to what they had originally.  And then the process repeats, indefinitely.  (See Figure \ref{world_loops}.)  If we observe (that is, consult an observable) with poor time resolution it will sample and receive contributions from many cycles of this motion.   It is an analogous process, in time, to how a spatial aperture samples a periodic process in space.   And here too we will have interference effects, leading to cancellations off resonance and a peaked spectrum.  

These considerations emphasize that the entities usually called, and considered to be, {\it virtual\/} particles (or states) can take on very tangible {\it reality}.   The quarks and gluons of QCD epitomize that phenomenon, but it is a much more general aspect of energy-time complementarity.   {\it Real virtuality\/} is the deep message of confinement.

\section{Back to QCD: Self-Adhesive Glue}

With that instructive model in mind, let us return to QCD.  In the model, the essence of the confinement mechanism was that the electric flux density associated with charge did not spread out, but rather stayed constant, as a consequence of Gauss' law applied to a one-dimensional distribution.   I three dimensions, of course, ordinary electric flux emanating from a charge source spreads out, with the fields $\propto 1/r^2$, and we have the familiar form of Coulomb's law.   QCD electric flux is quite different, however.   Since gluons themselves are not color neutral, they interact with one another directly.  In fact they strongly attract one another.  We might say that they are sticky, or self-adhesive.    Thus, it is seductive to imagine that they gather into tubes, leading us to dynamics that resembles 1+1 QED, with a constant force law.  

To keep things simple, in what follows  I will focus on the pure glue color $SU(3)$ variant of QCD, admitting quarks only as prescribed color sources, with no independent dynamics.  Real-world QCD of course contains some very light quarks.  Their existence complicates the discussion of flux tubes.  It brings in the possibility of ``string breaking'', whereby some of the energy associated with a  long flux tube gets invested into creating in  quark-antiquark pairs, allowing the tube to fragment.  In the pure glue theory, however, that doesn't occur.  

We have a conserved $Z_3$ triality quantum number, associated with the center of the gauge group $SU(3)$.  Color triplet sources emit a unit of triality flux, and anti-triplet sources absorb one.  The triality flux, being conserved, is subject to a Gauss law constraint, so that flux can't disappear, even at great distances.  Thus, we can plausibly anticipate that self-adhesive glue organizes itself into effectively one-dimensional quantized flux tubes, whose energy is approximately proportional to their length, as in 1+1 QED. 

We can motivate the ultra-qualitative idea of self-adhesive glue in several ways.  At a primitive level, two close-by octet gluons can combine into a singlet, lowering the field energy.   The basic phenomenon of asymptotic freedom is a more sophisticated, properly relativistic indication of flux concentration, and  it can be traced to attractive color paramagnetic forces between gluons \cite{af_para}.   The strong-coupling expansion of lattice QCD gives, in lowest order, simple line-like flux tubes \cite{wilson}.   Large $N$ expansions (for $SU(N)$) indicate the dominance of sheet-like space-time gluon configurations in the 't Hooft limit of finite $g^2/N$, $N\rightarrow \infty$ \cite{tH2}, which takes us beyond ordinary perturbation theory.   

\begin{figure}[hbt!]
    \centering
    \includegraphics[width=\columnwidth]{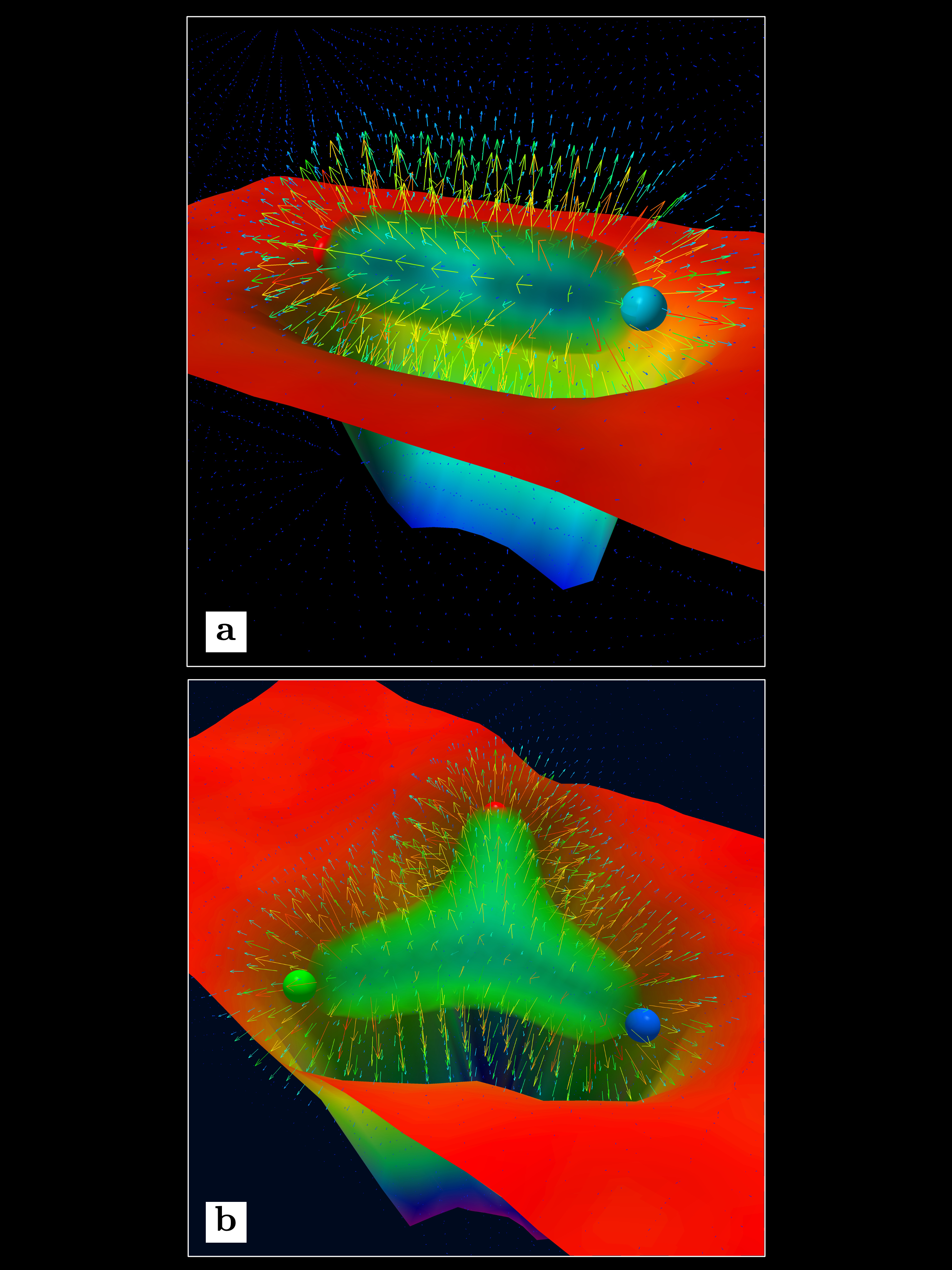} 
    \caption{Visual display of the field strength accompanying {\bf a:} separated color triplet and anti-triplet sources in pure glue $SU(3)$ and {\bf b:} three separated triplet sources, from \cite{website}.  The underlying data comes from calculations based on the full theory, with no uncontrolled approximations. }
\label{flux_tube_animation} 
\end{figure}

Of course, as Feynman liked to say, it is better to have one really good argument.  Fortunately, modern algorithms and supercomputers make it possible simply to sample the field configuration that arises in response to separated triplet and anti-triplet sources within a fully controlled calculation featuring small error bars \cite{leinweber, website}.  (See Figure \ref{flux_tube_animation}.)  Figure \ref{flux_tube_animation} is a picture worth thousands of words.


\section{Introducing Treeons}

While many aspects of the general picture outlined above hold in $SU(N)$ for any $N \geq 2$, some special features arise for $N=3$.  Because those features are central to the existence and structure of baryons, and thus of matter as we know it (and consist of) they are basic ingredients in the theoretical description of the physical world.   

Beautiful calculations \cite{japanese_tubes} and animations by Leinweber \cite{leinweber_treeon} and his collaborators show that the three unit flux tubes that emanate from well-separated triplet sources in triangular arrangements generally come together in a small hub. It is plausible, though not yet fully established, that these hubs have characteristic properties that appear in diverse circumstances.  Let's try to take that hypothesis seriously.   In view of the association of ``tre'' ``tres'', and of course ``three'' with the number 3, and the emergence of tree-like flux structures when we bring in more complex distributions of sources and anti-sources, let us call them {\it treeons\/} (and anti-treeons).   

If we idealize QCD flux-tubes as thin strings, and suppose that the energy of a flux configuration is exactly proportional to the length of string involved, then the energetics of prescribed color triplet and anti-triplet sources defines a physically motivated mathematical problem.  It generalizes the classic Steiner problem of joining prescribed points by a network of lines with minimal total length.   The rules of this modified problem are as follows.  We are given $M$ points representing color sources and $N$ points, with $M-N \equiv 0$ modulo 3, all with fixed positions.  To this configuration we can add $S$ treeon points and $T$ anti-treeon points,  with $3(S-T) = M - N$.  Single directed lines emanate from the sources and go into the anti-sources, while triples of directed lines go into treeons and emanate from anti-treeons.  The lines have energies proportional to their length.   The basic energy minimization problem is to choose $S, T$ and the positions of the treeons and anti-treeons in such a way as to minimize the total length of the network, subject to those rules.   

\begin{figure}[hbt!]
    \centering
    \includegraphics[width=\columnwidth]{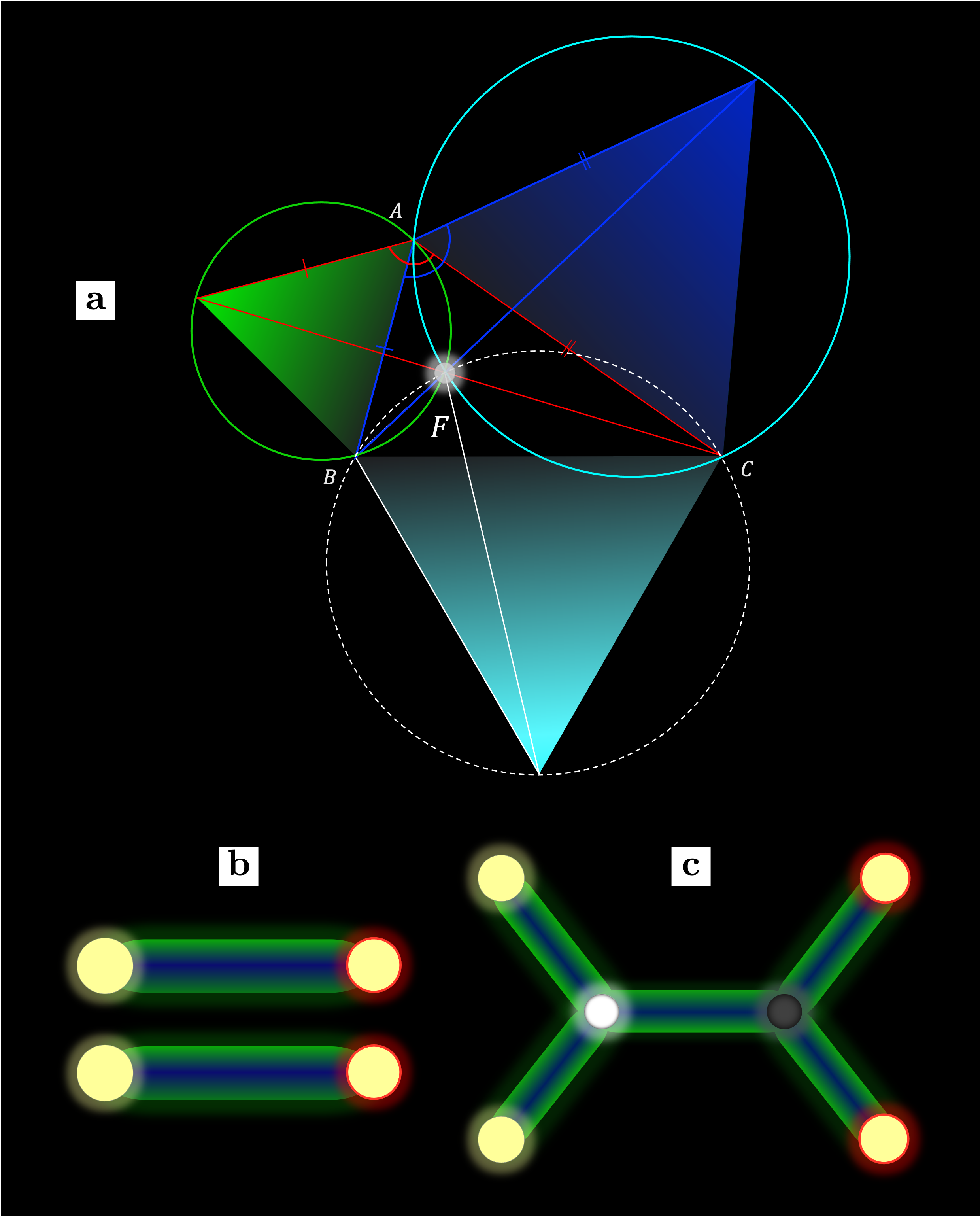} 
    \caption{{\bf a:} Geometric construction of the Fermat point of a triangle: It is the intersection of circles circumscribed about equilateral triangles erected on its side.  When the large angle of an obtuse triangle exceeds $2\pi/3$, we have a ``degenerated'' Fermat point coincident with that angle's vertex.  {\bf b, c:} With two triplet and two anti-triplet sources, it can become favorable to create a treeon-anti-treeon pair, enabling transition from two-meson and tetraquark flux distributions.}
\label{fermat} 
\end{figure}

The simplest non-trivial case with $M = 3, N =0$, is a classic problem in Euclidean geometry, solved by Torricelli and Fermat.   In simple terms, it asks for the location of the point that can be joined to the vertices of a triangle with minimal total length.    The solution, known as the Fermat point, though complicated to express in Cartesian coordinates, has an elegant geometric construction, being the intersection of circles circumscribed around equilateral triangles erected on the triangle's sides.   For our purposes, the most important property of the Fermat point, in non-degenerate cases,  is that the lines which join it to the vertices of the triangle make equal angles $2\pi /3$ with one another. (See Figure \ref{fermat}a.)
Note however that if the largest angle in the triangle exceeds $2\pi /3$, we get the minimal length with a point that coincides with the vertex of that angle. In this case a degenerated treeon coincides with a source, and connects to it by a flux line of zero length.   

This minimization problem, taken in full generality, is NP complete, and there is a lot to say about it as a mathematical problem.  For present purposes, an important general result that in minimal-length solutions the lines associated with non-degenerate treeons and anti-treeons (i.e., those that do not lie on sources or anti-sources) are coplanar and make $2 \pi /3$ angles with one another.  With that result in hand, it is an easy geometric exercise to analyze the case $M = N =2$ when the sources are two adjacent vertices of a rectangle and the anti-sources are the other two vertices. (See Figure \ref{fermat}b, d.)  As depicted, there are two candidate solutions: a treeon-free ``two meson'' configuration where the lines run along two sides of the rectangle from sources to anti-sources, and a ``tetraquark'' configuration containing a treeon and an anti-treeon.   Which of those  is more favorable depends on the aspect ratio of the rectangle; one calculates that they cross over when the ratio of the source--source sides to the source--anti-source sides ratio is $1/\sqrt 3$.  

How well does this idealized model reflect actual (pure glue) color gauge theory?  That question has received significant attention over the years, but does not appear to be fully resolved \cite{Y_skeptics}.   Clear tube-like structures emerge only as the sources are quite distant from one another, and therefore necessarily involve states of large energy, which are challenging to access numerically.   If we can rise to that challenge, several fundamental, conceptually interesting questions arise, including:

\begin{itemize} 

\item Are flux tubes carrying unit triality the preferred way to transport $Z_3$ flux over large distances?  

This can be addressed by putting in sources with different values of triality, {\it e. g}., color sextets.  (Work with diquarks \cite{diquarks} is very relevant here.)  According to this (very plausible) hypothesis, the flux tube connected far-separated source-anti-source pairs of this kind will have the same properties, after a healing length, as the flux tube connected anti-triplet-triplet pairs.  

\item Are treeons a robust emergent structure?  

This can be addressed by checking whether the Fermat equiangular condition obtains in a wide variety of geometries.  

\item If so, what are their properties?

Since treeons arise in a non-relativistic framework, they have both a gap (which could be negative) and a kinetic mass.   The gap can be measured in the energy of static configurations (sources on spatially fixed world-lines) while the kinetic mass appears in the effective action for curved source world-lines. 

\end{itemize}

All these questions remain interesting in 2+1 dimensions.  They should be considerably easier to answer there, both for the obvious reason that there are fewer degrees of freedom to keep track of, and for the more specific reason that flux has less motivation to swell out sub-asymptotically.  

\section{Concept of Flux Channeling}

As already mentioned, numerical calculation of flux-tube and treeon properties are challenging.   Due to end effects near the sources we must keep them far separated.  That implies, other things being equal, that much of the calculation will be wasted on reproducing the properties of ``empty'' space, which of course is full of fluctuating fields.  

\begin{figure}[hbt!]
    \centering
    \includegraphics[width=\columnwidth]{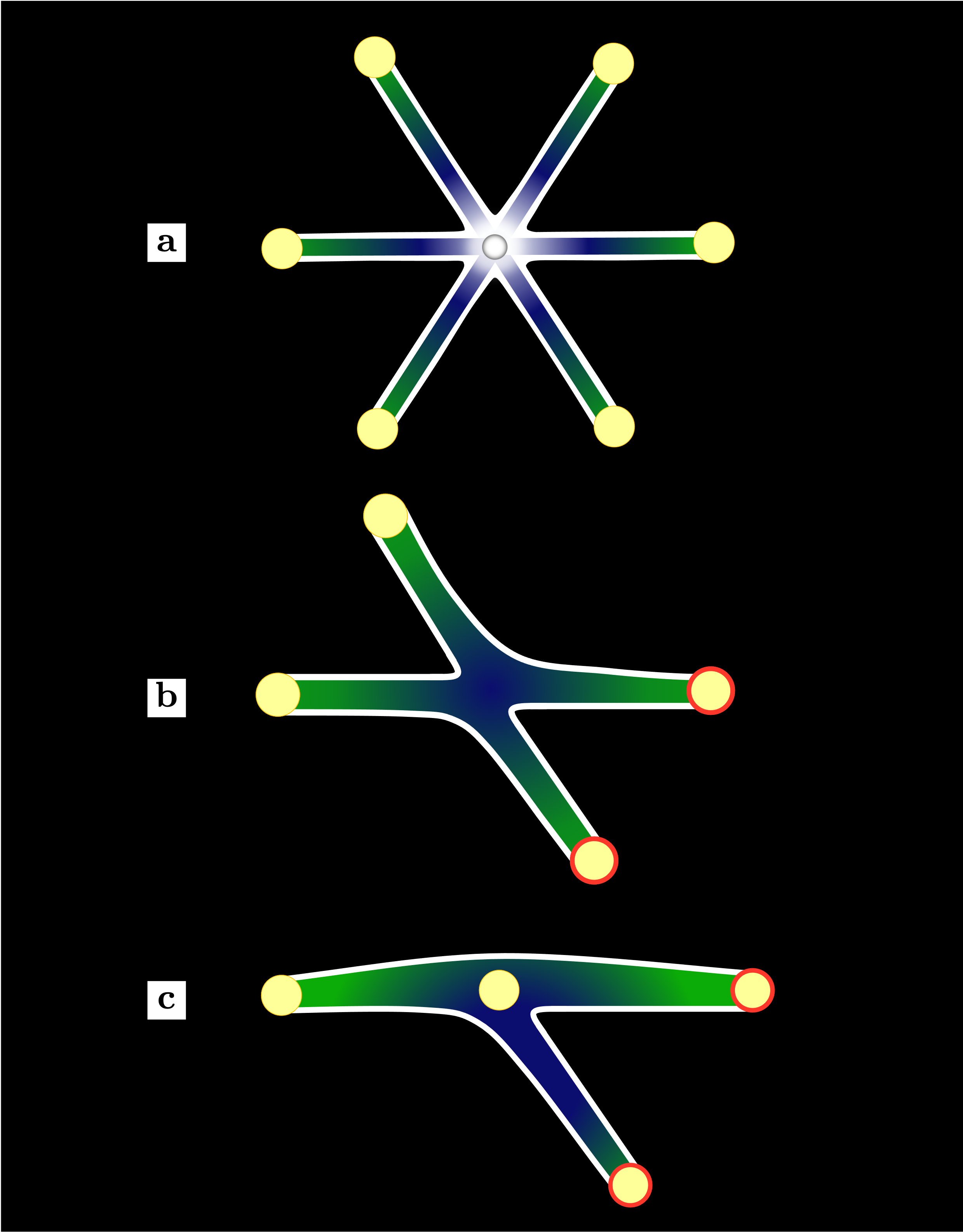} 
    \caption{By channeling flux, we can access interesting interactions efficiently.  Here we see channeling configurations that facilitate {\bf (a)} treeon-treeon interactions, {\bf (b)} flux tube-flux tube interactions, and {\bf (c)} flux-tube-triplet interactions.} 
\label{channeling_configurations} 
\end{figure}

We can relieve those geometric deficiencies, and open up some otherwise inaccessible dynamics to scrutiny, by artificially confining the fields to channels.  This can be one in a gauge-invariant way, by allowing the coupling constant to become very small outside the channels.    In this way we can access treeon-treeon interactions, tube-tube interactions, and tube-quark interactions, for example. (See Figure \ref{channeling_configurations}.) We can even study isolated quarks and the flux emerging from them.  Flux tubes need not extended indefinitely, because with 
\begin{equation}\label{channel}
{\cal L} \sim \frac{1}{2g^2(r)} E^2 \, + q \delta^3(x)
\end{equation}
we have the modified Gauss law
\begin{equation}
4\pi q = \frac{1}{g^2(R)} \Phi_E (R)
\end{equation}
which allows the flux $\Phi$ to die together with $g^2$.  

\section{Flux Channeling in Materials}

Looking at the material Maxwell Lagrangian density 
\begin{equation}
{\cal L} ~=~ \frac{\epsilon(x)}{2} \, E^2 \ -\ \frac{1}{2\mu (x)} \, B^2
\end{equation}
indicates, by comparison with Eqn.\,(\ref{channel}), that $\epsilon(x) \rightarrow \infty, \mu(x) \rightarrow 0$ correspond to ``dead zones'' for electric and magnetic fields, respectively.  The latter is particularly interesting, because it corresponds the perfect diamagnetism of superconductors.  Thus, magnetic channeling in superconductors can be implemented by drilling holes.   This opens up interesting possibilities for transporting and sculpting magnetic flux.  

Here too there is much more to say.  Issues around treeons and flux channeling, touched on here, are under active investigation \cite{mit_group}. 

The future beckons ... 

\section{A Brief Futurology of QCD}

My crystal ball brings up several images: 

\subsection{A Tool Supporting Experimental Exploration}

\begin{figure}[hbt!]
    \centering
    \includegraphics[width=\columnwidth]{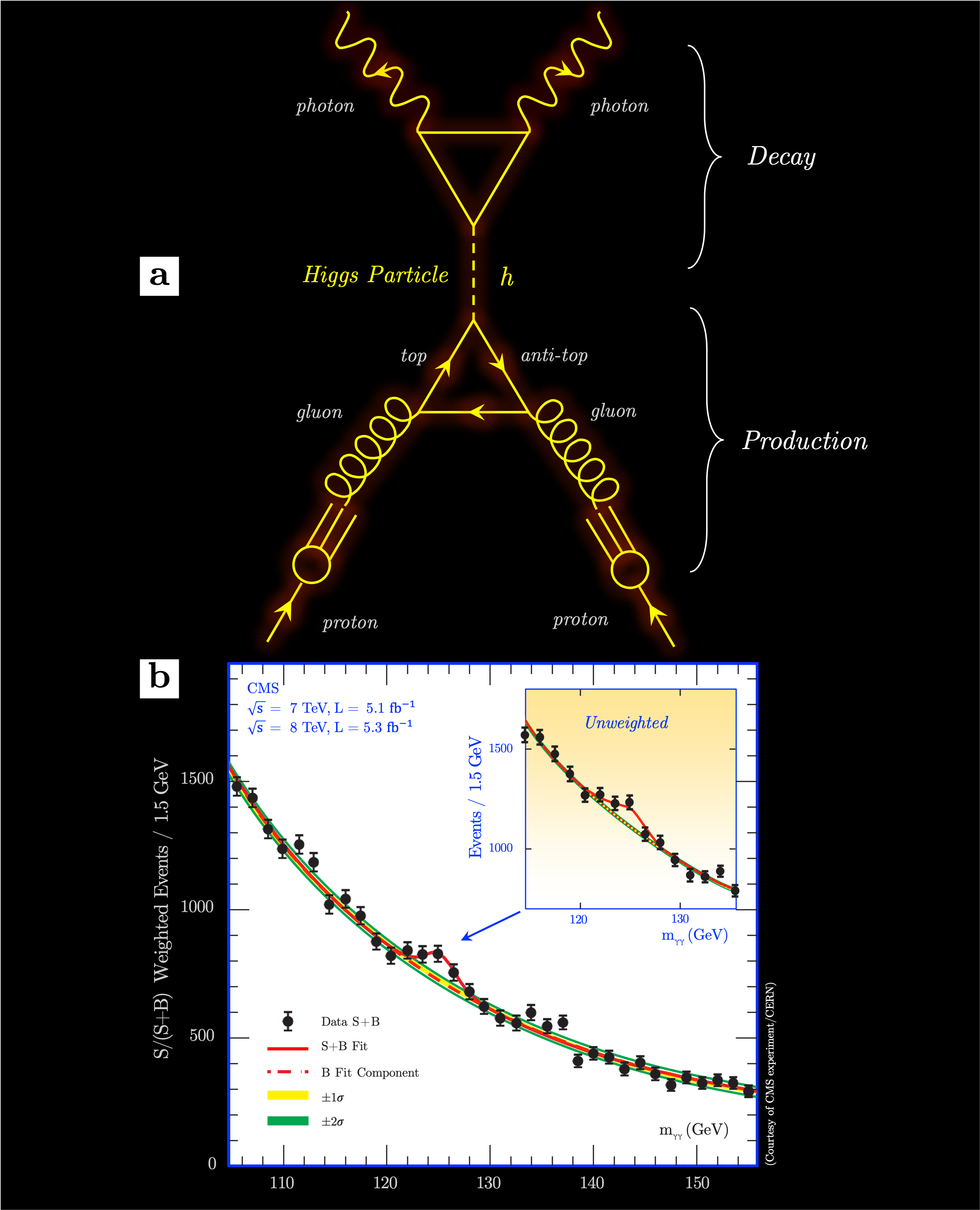} 
    \caption{{\bf a:} Hybrid Feynman graph description of the process underlying the original Higgs particle discovery. Protons are idealized as gluon sources (!) and highly virtual top quarks power the production mechanism.  {\bf b:} Experimental data (CMS collaboration) that constituted the discovery: a statistical enhancement in the $\gamma \gamma$ invariant mass distribution. Both the search strategy the experiment's interpretation would have been inconceivable (literally) without strong guidance from QCD.}  
 \label{higgs_discovery} 
\end{figure}

QCD played a crucial role -- two crucial roles, actually -- in making the discovery of the Higgs particle possible.  One was to show how the Higgs particle could be produced at a hadron accelerator, and to give quantitative estimates of the rate.  This involves realizing that protons contain a lot of gluons, and then invokes the beautiful idea \cite{fw_H_glue} that gluons communicate with Higgs particles through virtual top quarks  (See Figure \ref{higgs_discovery}.)  Since the Higgs particle is terribly unstable, though, it is not enough to produce it.  One must find distinctive signatures of its fleeting existence.  QCD's second role was to support quantitative estimates of ``everything else'' that might contribute.  As it happened, the initial discovery was statistical -- a small enhancement in the production of di-photons within a small mass range.   Evidently, knowing what to look for, {\it quantitatively}, was crucial both to the design of the experiment and to its interpretation   The discovery of low-energy supersymmetry, should it ever occur, will involve many stories of this kind.

QCD will continue to guide the design and interpretation of exploratory experiments in fundamental physics.   Here I should mention not only experiments at high energy, but also probes of fundamental symmetry, including the study of rare decays (including, let us hope, proton decay) and ``forbidden'' properties (including, let us hope, neutron and other nuclear electric  dipole moments).

\subsection{A Tool for Astrophysics and Nuclear Technology}

There are important astrophysical problems that provide major challenges and opportunities for QCD.  Notably, the emergence of gravitational wave and multi-messenger astronomy will provide new windows into the deep structure of neutron stars.   Neutron stars, of course, are the ultimate nuclei.  Their theory should, and eventually will, be brought to the same level as that we have achieved for ordinary stars.   What is the equation of state?  Are there qualitative changes and discontinuities approaching the core?   Despite knowing the governing equations, we don't have convincing answers even to such basic questions.  

Other sorts of extreme nuclear physics comes into play in the violent events around supernova explosion (of several kinds) and neutron star mergers.    Elucidating all this is part of the project of understanding the QCD phase diagram and the intellectually adjacent dynamics.  

In the charming and brilliant science-fiction novel {\it Dragon's Egg} \cite{forward}, Robert Forward extrapolates from the complex and poorly understood physics of neutron star crusts to the emergence of a new form of intelligence.   These creatures, whose thought-processes take place on nuclear rather than chemical time-scales, are quick-witted indeed.  Following exposure to human culture, they rapidly absorb and then outstrip it.  Forward's ``application'' of nuclear chemistry is probably far-fetched, but it serves here as a proxy for the possibility that profound understanding and control of nuclear behavior plausibly will support important applications -- QCD-based  clocks, batteries, and lasers, for example, or fuel.  

\subsection{A Pointer to New Realities}

QCD brings quantum theory, special relativity, and symmetry together in an intense union.   Particle creation, vacuum polarization, field topology, gauge symmetry, anomalies, renormalization, spontaneous symmetry braking, exotic superconductivity -- not to mention confinement and asymptotic freedom -- most of which are small, subtle, or isolated effects elsewhere, come together and are woven into the fabric of QCD.   Mixing metaphors: QCD is where the rubber of deep quantum theory meets the road of physical reality.  As such, it is a domain in which the challenge of understanding natural phenomena can stimulate hard thinking and new methodologies.  Modern lattice gauge theory \cite{lgt}  which brings in sophisticated sampling methods, exquisite algorithms and, recently, machine learning, is an outstanding example.  So is the exploration of heavy ion collisions \cite{heavy_ion}, where the AdS-CFT connection \cite{maldacena} has found perhaps its most impressive physical application \cite{son}, as is the elucidation of confinement in highly supersymmetric gauge theories \cite{seiberg-witten}.   Our earlier discussion of energy-time complementarity fits in here, too.

\begin{figure}[hbt!]
    \centering
    \includegraphics[width=\columnwidth]{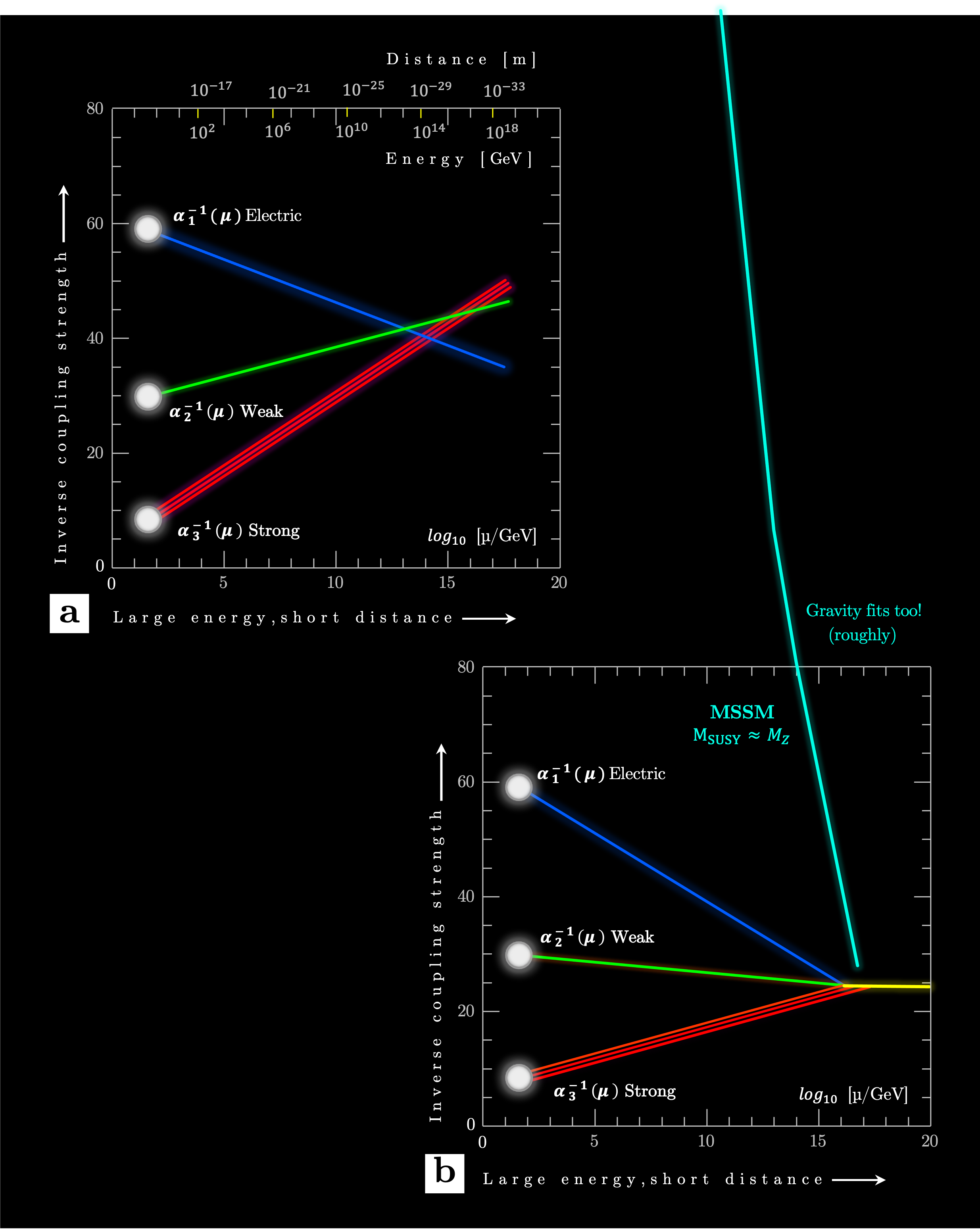} 
    \caption{Schematic indication of the extrapolation of gauge theory coupling without {\bf (a)} \cite{gqw} and with {\bf (b)} \cite{drw} contributions from low-energy supersymmetry.  The qualitative and, in the latter case, semi-quantitative indication of unification, also bringing in gravity, encourages us to take these ideas seriously (and literally). } 
\label{coupling_unification} 
\end{figure}

Finally: As a secure, beautiful, and mathematically rich part of the fundamental description of Nature, QCD will continue to support and inspire the quest for more comprehensive theories, and to guide us in deducing their consequences. The iconic unification of couplings calculation \cite{gqw}\cite{drw}  is an outstanding example. (See Figure \ref{coupling_unification}.)  Axions are another, whose time may be at hand \cite{alpha}.  

\bigskip

\bigskip

\section{Postscript: Erice 1973}

For me, 1973 was a watershed in more ways than one.  I started it as a 21-year-old bachelor graduate student at Princeton, recently converted from mathematics to physics.  I was recovering from two years ``in the wilderness'' after graduating from the University of Chicago, during which  I lacked clear direction. 

Things were looking up, though:  In the summer of 1972 I'd met Betsy Devine, and we'd become an item.   Also I'd started work to work in earnest with David Gross.  As 1972 closed, I was well into applying the renormalization group to non-abelian gauge theories and calculating their $\beta$ functions.   

One year later, by the end of 1973, I'd married Betsy.  (We recently celebrated our golden anniversary.)  The calculations had gone well, too, and David and I had written the papers that launched modern QCD.  

The Ettore Majorana Summer School on High Energy Physics, which I attended as a student, was the third formative event in my 1973.  Until then my exposure to the community of research physics had been extremely limited, in ways that I wasn't even aware of.   Basically, apart from David Gross and (to a lesser extent) Sidney Coleman, I had mostly encountered other scientists as teachers and authority figures, as opposed to human beings.   The special atmosphere at Erice, where lecturers and hearers inhabited the same small town and shared meals, and where questions were actively encouraged, gave me new perspectives.  I got to see and hear how leading professional scientists think and talk with one another -- their special ``high-level language'', so to speak, as opposed to machine code.   I had known Sidney Coleman during his sabbatical leave at Princeton in the preceding months, where we frequently talked physics, but at Erice our friendship, nurtured by his wit and humanity, deepened.  Shelley Glashow was another lecturer, and showed me a different style, also very attractive, full of joy and enthusiasm; Sam Ting -- not yet as famous as he soon would be -- was impressive, bordering on scary, with his intensity and drive.  Later, over many years they all became my friends and colleagues, building on the foundations laid at Erice.  

I also got to spend a few magical hours with I. I. Rabi,  in a series of one-on-one lunches.  I think he was happy to speak with a fellow New Yorker.  I wrote a piece about this in the {\it Wall Street Journal\/}, which I'll now briefly quote:

`` ...  I was full of theoretical ideas and quasi-philosophical speculations. Rabi pressed me -- gently, with a twinkle in his eye, yet relentlessly -- to
describe their concrete meaning ... Several times, an exasperated Rabi had asked me, in response to my speculations, `OK, but what am I
supposed to do when I come in to the lab in the morning?'  I was tempted to say `That’s your problem,'
but of course I bit my tongue. Eventually I realized what he really meant: Fully worked-out answers to
good scientific questions should include solid experimental prospects.''

Antonio Zichichi, vigorously  presiding over it all, was the epicenter of my, and everyone's, Erice experience.  Back in 1973  I couldn't properly appreciate the magnitude of his achievement.  Still less did I  have any thought of ever attempting to do anything remotely similar.  But now, decades later, after mounting  -- together with Antti Niemi, Wu Biao and others -- a series of Quantum Connections schools that try to reproduce the same spirit, I am properly awed.

\bigskip

\bigskip

\section*{Acknowledgements} This work is supported by the U.S. Department of Energy under grant Contract  Number DE-SC0012567, by the European Research Council under grant 742104, and by the Swedish Research Council under Contract No. 335-2014-7424. I would like to thank Juliana Baena for many helpful suggestions and for greatly improving the figures.

\newpage

\section*{Figures}


\begin{thebibliography}{99}
\bibitem{aps_collection} A curated selection of papers to celebrate 50 years of QCD is available at  journals.aps.org/collections/50-years-QCD .
\end{thebibliography}
\end{document}